# Probing the relativistic mean field effective *NN* forces for fusion of spherical colliding nuclei within a dynamical model

Maria V. Chushnyakova[1], Igor I. Gontchar[2], Natalya A. Khmyrova[2]

[1]*Physics Department, Omsk State Technical University, 644050 Omsk, Russia*
[2]*Physics and Chemistry Department, Omsk State Transport University, 644046 Omsk, Russia*

For the first time, the precise data on the above barrier fusion (capture) cross-sections for the reactions involving spherical colliding nuclei are quantitatively analyzed using the relativistic mean field effective interaction. The parameter sets NL1, NL2, NL3, and HS are employed. The analysis is performed within the framework of the fluctuation-dissipation model with surface friction based on the double-folding approach for the nucleus-nucleus potential. The effective interactions, as well as the resulting potentials, are confronted with the ones obtained using the M3Y *NN* forces. Of the four studied *NN* interactions, the Coulomb barrier appears for the nucleus-nucleus potentials corresponding to the NL2 and HS parameter sets when the exchange forces are added to the effective interaction. The heights and radii of the barriers obtained using these two parameter sets are very close to each other. The NL2 potential is used for analyzing the fusion cross-section data for five reactions. The results of dynamical calculations are in good agreement with the experimental data.

* Corresponding author.
*E-mail address:* maria.chushnyakova@gmail.com (M. Chushnyakova).

The Relativistic Mean Field theory (RMF) has proven to describe successfully static properties of nuclei like binding energies and Coulomb form-factors [1–3]. Yet there are plenty of precise data (with the typical error of 1%) on fusion (capture) of complex nuclei [4,5] which have been studied in detail using the double-folding approach with M3Y effective *NN* forces [6,7]. Thus, it seems logical to probe describing the same data using the effective *NN* forces resulting from the RMF approach.

The effective RMF *NN* interaction reads [8]

$$v_{NN}(r) = \frac{g_\omega^2}{4\pi}\frac{\exp(-m_\omega r)}{r} + \frac{g_\rho^2}{4\pi}\frac{\exp(-m_\rho r)}{r} - \frac{g_\sigma^2}{4\pi}\frac{\exp(-m_\sigma r)}{r} + \frac{g_2^2}{4\pi}r\exp(-2m_\sigma r) + \frac{g_3^2}{4\pi}\frac{\exp(-3m_\sigma r)}{r} - J_{00}\delta(r). \quad (1)$$

The values of the meson masses $m_\omega$, $m_\rho$, and $m_\sigma$, as well as the couplings $g_\omega$, $g_\rho$, $g_\sigma$, $g_2$, and $g_3$, obtained from the fits of the nuclear static properties are presented in Table I with the corresponding references.

TABLE I. The values of meson masses and corresponding couplings obtained from the fit of static properties of nuclei with the proper references

|  | NL1 [1,2] | NL2 [2] | NL3 [8,9] | HS [10–12] |
|---|---|---|---|---|
| $m_\omega$ (MeV) | 795.359 | 780.0 | 782.501 | 783 |
| $m_\rho$ (MeV) | 763.0 | 763.0 | 763.000 | 770 |
| $m_\sigma$ (MeV) | 492.25 | 504.89 | 508.194 | 520 |
| $g_\omega$ | 13.285 | 11.493 | 12.868 | 13.8 |
| $g_\rho$ | 4.975 | 5.507 | 4.474 | 8.08 |
| $g_\sigma$ | 10.138 | 9.111 | 10.271 | 10.47 |
| $g_2$ (fm$^{-1}$) | -12.172 | -2.304 | -10.431 | 0 |
| $g_3$ | -36.265 | 13.783 | -28.885 | 0 |
| $J_{00}$ (MeV fm$^{-3}$) | -592 (Paris) | -592 (Paris) | -592 (Paris) | -276 (Reid) |

The structure of the effective RMF *NN* interaction is similar to that of the well-known M3Y interaction:

$$v_{NN}(r) = G_1 \frac{\exp(-r/r_1)}{r/r_1} - G_2 \frac{\exp(-r/r_2)}{r/r_2} - J_{00}\delta(r). \quad (2)$$

In the literature one finds two parameter sets for this interaction: $G_1$=7999 MeV, $G_2$=2134 MeV, $J_{00}$= 276 MeV·fm$^3$ (the so-called Reid forces [13], M3Y_R) and $G_1$=11062 MeV, $G_2$=2537.5 MeV, $J_{00}$= 592 MeV·fm$^3$ (Paris forces [14], M3Y_P). The values of the radius parameters are same for both sets: $r_1$= 0.25 fm, $r_2$=0.40 fm. Note that we omit the energy dependence of the interaction since it is insignificant for the present consideration. The M3Y and RMF *NN* forces are compared in Fig. 1. One sees that all six profiles look rather similar implying similarity of the nucleus-nucleus interaction resulting from these *NN*-forces.

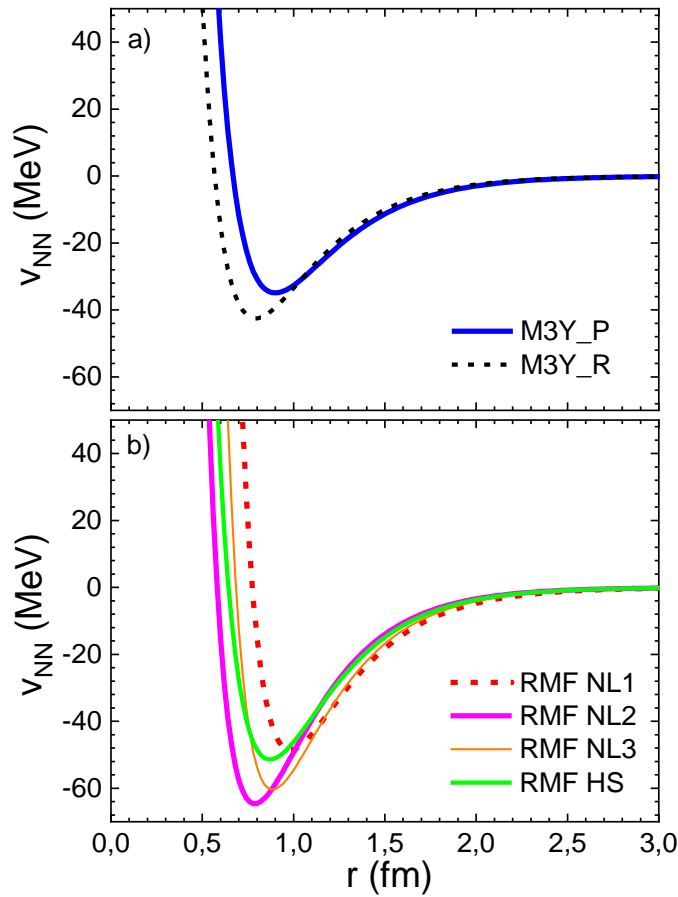

FIG. 1. Effective nucleon-nucleon interaction: (a) M3Y [Eq. (2) except the delta-function term] and (b) RMF [Eq. (1) except the delta-function term].

We evaluate the Strong nucleus-nucleus potential (Snn-potential, $U_n$) by means of the double-folding approach:

$$U_n(R) = \int d\vec{r}_P \int d\vec{r}_T \rho_P(\vec{r}_P) v_{NN}(|\vec{R} + \vec{r}_T - \vec{r}_P|)\rho_T(\vec{r}_T). \quad (3)$$

Here $\vec{R}$ denotes the vector joining the centers of mass of the colliding nuclei, $\vec{r}_P$ and $\vec{r}_T$ are the radius vectors of two interacting points of the projectile and target, respectively. The figure for the geometry can be found in [15,16]. The Coulomb interaction is calculated using the double-folding formula analogous to (3). The nuclear matter and the charge densities required for these calculations are taken from [7]. They have been obtained using the Skyrme-Hartree-Fock approach with the tensor forces as described in [17] with the SKX parametrization of Ref. [18]. The resulting Snn-potentials and total interaction potentials (including the Coulomb term) calculated for the s-wave in reaction $^{12}$C+$^{92}$Zr are displayed in Fig. 2. The Snn-potentials in panel (a) corresponding to RMF without the delta-function term (i.e. NL1_0, NL2_0, and NL3_0) do not show any decrease as *R* becomes smaller. These Snn-potentials result in the total potentials $U_{tot}(R)$ without barriers in panel (b). This means that no description of the heavy-ion fusion (capture) data can be obtained with these potentials. In Fig. 2a one sees that the potential $U_n$ obtained with NL2_0 parameter set increases with the decrease of $R$ significantly slower than the two others (NL1_0 and NL3_0).

For the comparison, we include in Fig. 2 the M3Y potentials with Reid (M3Y_R) and Paris (M3Y_P) parameters. The corresponding curves look very similar in both panels.

As the next step, we evaluate the RMF $U_n$ potentials including the values of $J_{00}$ indicated in Table I. In the cases of NL1 and NL3 parameter sets, this does not help: the corresponding total interaction potentials still do not possess barriers. However, the potential $U_n$(NL2_P) decreases similar to $U_n$(M3Y) (Fig. 2a) resulting in the barrier in Fig. 2b. The shape and height of this barrier are remarkably close to that of the M3Y barriers. The HS-forces of Ref. [11] with $J_{00}$= 276 MeV·fm$^3$ result in the nucleus-nucleus potential which also decreases as the center of mass distance does so (see Fig. 2a). Thus, the total interaction potential $U_{tot}$(HS_R) possesses a barrier whose height and position are very much similar to those of $U_{tot}$(NL2_P) (see Fig. 2b).

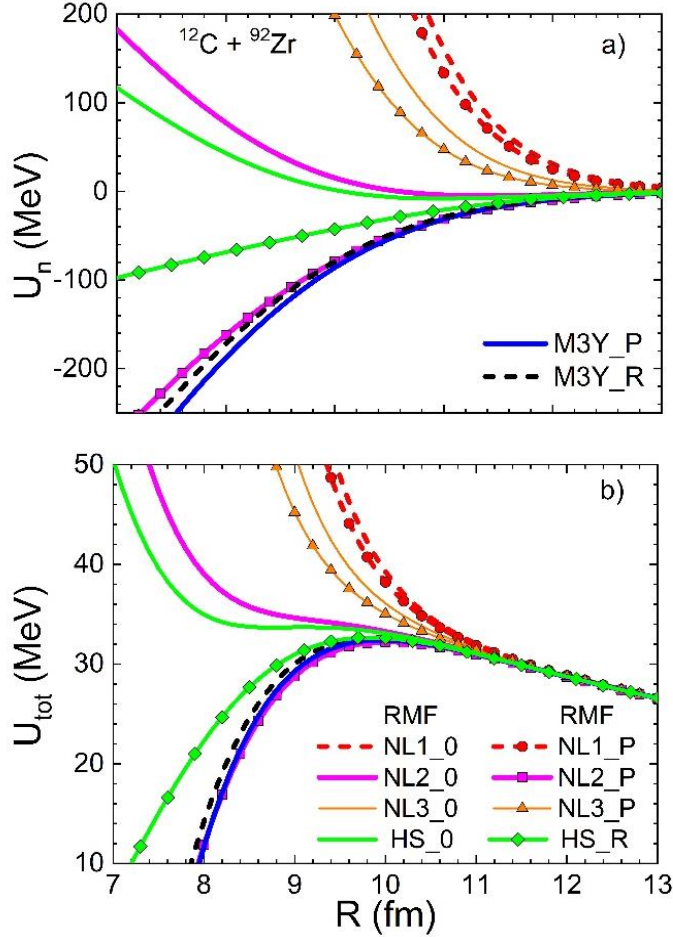

FIG. 2. SnnP (a) and total nucleus-nucleus interaction potential (b) versus the center-of-mass distance for the reaction $^{12}$C+$^{92}$Zr. Potentials NL1_0, NL2_0, NL3_0, HS_0 are evaluated according to Eq. (1) without the delta-function term whereas potentials NL1_P, NL2_P, NL3_P, HS_R are calculated with the values of $J_{00}$ indicated in Table I.

Let us now trying to understand the behavior of the Snn-potentials calculated on the basis of the RMF approach considering the corresponding effective *NN* forces in Fig. 1. Comparing the RMF curves in Fig. 1 with the M3Y curves, one could expect that all the Snn-potentials based on the RMF would decrease with the decrease of $R$ producing a barrier. Indeed, the forces v$_{NN}$(RMF) look similar to v$_{NN}$(M3Y), moreover, v$_{NN}$(RMF) are even deeper.

To figure out this apparent contradiction we plot in Fig. 3 all six v$_{NN}$ potentials at small values of the nucleon-nucleon distance. One sees that v$_{NN}$(NL1) and v$_{NN}$(NL3) increase much faster as $r$ decreases than the four others do. Moreover, the $r$-dependences of the v$_{NN}$(NL2) and v$_{NN}$(HS) become very close to the one of the v$_{NN}$(M3Y_P) at small values of $r$: these tree potentials form a close group of thick solid curves in Fig. 3. Thus, using these three v$_{NN}$, it is natural to expect barriers with similar heights and radii.

The repulsion in v$_{NN}$(NL1) and v$_{NN}$(NL3) becomes strong at too large distances between the nucleons in comparison with the four remaining kinds of v$_{NN}$. By our opinion, this observation solves the puzzle.

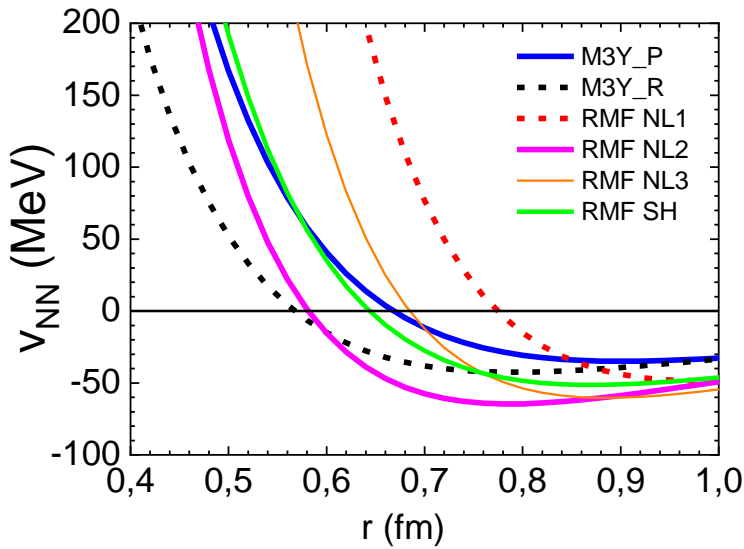

FIG. 3. Six considered effective nucleon-nucleon interactions without delta-function terms: more detail view in the domain of small $r$-values.

The nucleus-nucleus interaction energy resulting from the RMF approach has been calculated earlier in [10–12]. The NN forces from Ref. [11] are shown in Figs.1 and 3, the resulting barriers are presented in Table II. Our calculations approximately reproduce the barrier for $^{14}$C+$^{208}$Pb of Fig. 2 of that work. In fact, we calculated the $^{12}$C+$^{208}$Pb system and obtained $U_{B0}$(NL2_P)=58.3 MeV and $U_{B0}$ (HS_R)=58.9 MeV.

A comparison of our results with those of [10] is problematic: the approaches are too different. In particular, we use the SKX Hartree-Fock densities whereas in [10] the relativistic Hartree-Bogoliubov densities are used. For the NN forces, in [10] the delta-function approximation is used, whereas we apply the RMF $v_{NN}$ with the finite interaction radii.

It looks like our result, i.e. the absence of the barrier with the NL3 forces, is in contradiction with the results of [12] where the fusion cross sections are calculated using this NN interaction. One source of the apparent disagreement can be the different densities used in [12] and in the present work. Namely, in [12] the densities resulting from the NL3 interaction are applied whereas we use the SKX Hartree-Fock densities. Another possible reason is that the different reactions are used in [12] and in our work. The question about the apparent disagreement probably calls for further study.

We now go over to the comparison with the experiment. For this aim we choose the high precision data on the above-barrier portions of the fusion (capture) cross sections for the reactions $^{16}$O+$^{92}$Zr and $^{28}$Si+$^{92}$Zr [19], $^{16}$O+$^{144}$Sm [20], $^{16}$O+$^{208}$Pb [21], $^{12}$C+$^{144}$Sm [22] (the data are often taken from the database [23]). Note that typical experimental errors in [19–21] are between 0.5-1% whereas in [22] they are about 5%.

We calculate the cross sections within the framework of the fluctuation-dissipation trajectory model with surface friction of Refs. [6,7,24]. Since the model is described and tested in detail in those papers, we give here only short overview of it.

The physical picture of the model is similar to that of Refs. [25]: the fictitious Brownian particle with the reduced mass wanders being affected by the conservative, dissipative, and random (fluctuating) forces. We study the process at the energies well exceeding the Coulomb barrier, therefore the quantum effects like tunneling and channels coupling are not accounted for. In the reactions considered only the spherical target nuclei are involved. They are rather stiff due to at least one (proton or neutron) closed shell. Therefore, only one degree of freedom corresponding to the radial motion is accounted for. The motion of the Brownian particle is described by the dimensionless coordinate $q$ which is proportional to the distance between the centers of the projectile and target nuclei $R$. In [26] it was checked that accounting for the orbital degree of freedom could be ignored since it influenced the cross sections within the statistical errors of the modeling (typically 1%).

In [27] it was shown that the memory effects appear in the heavy ion collision process only near the contact configuration. We never reach this configuration in our modeling. That is why we use the stochastic Langevin-type equations with the white noise and instant dissipation:

$$dp = (F_U + F_{cen} + F_D)dt + \sqrt{2D}\, dW, \qquad (4)$$
$$dq = pdt/m_q, \qquad (5)$$

$$F_U = -dU/dq, \tag{6}$$

$$F_{cen} = \frac{\hbar^2 L^2}{m_q q^3}, \tag{7}$$

$$F_D = -\frac{p}{m_q} K_R \left(\frac{dU_n}{dq}\right)^2, \tag{8}$$

$$D = \theta K_R \left(\frac{dU_n}{dq}\right)^2. \tag{9}$$

Here $p$ denotes the linear momentum corresponding to the radial motion; $F_U$, $F_{cen}$, and $F_D$ are the conservative, centrifugal, and dissipative forces, respectively. The latter is related to the Snn-potential via the surface friction formula (8) [28,29]. $L$ is the projection of the orbital angular momentum onto the axis perpendicular to the reaction plane; $m_q$ is the inertia parameter; $K_R$ stands for the dissipation strength coefficient; $D$ denotes the diffusion coefficient which is proportional to the temperature $\theta$. The random force is proportional to the increment $dW$ of the Wiener process $W$; this increment possesses zero average and variance equal to $dt$. Equations (4), (5) are solved numerically using the Runge-Kutta method (see details in [24,30]).

The capture cross sections are calculated using the standard quantum mechanical formula (see e.g. [31])

$$\sigma_{th} = \frac{\pi \hbar^2}{2 m_R E_{c.m.}} \sum_{L=0}^{L_{max}} (2L+1) T_L. \tag{10}$$

Here $E_{c.m.}$ is the collision energy; $m_R = m_n A_P A_T / (A_P + A_T)$ includes the nucleon mass $m_n$ and the mass numbers of the projectile ($A_P$) and target ($A_T$) nuclei; $L_{max}$ is the maximal angular momentum above which the transmission coefficient $T_L$ becomes small enough. The transmission coefficient appears as the result of the dynamical modeling described above.

To compare the calculated capture (fusion) cross-sections $\sigma_{th}$ with the experimental ones $\sigma_{exp}$, we calculate $\sigma_{th}$, varying the value of the dissipation strength coefficient $K_R$ in Eqs. (8), (9) for a given reaction similar to Refs. [7,24]. At each value of $K_R$ the value

$$\chi^2 = \frac{1}{v} \sum_{i=1}^{v} \left(\frac{\sigma_{ith} - \sigma_{iexp}}{\Delta \sigma_{iexp}}\right)^2 \tag{11}$$

is calculated. Here $\sigma_{ith}$ corresponds to the particular value of $E_{c.m.\,i}$ whereas $\sigma_{iexp}$ and $\Delta\sigma_{iexp}$ are the experimental value of the cross section and its error at the same value of the collision energy. We define the optimum value of the dissipation strength, $K_{Rm}$, searching for the minimum value of $\chi^2$, $\chi_m^2$. In Table II, the values of $K_{Rm}$, $\chi_m^2$, and $U_{B0}$ (the height of the Coulomb barrier at zero angular momentum) resulting from this study are compared with those obtained in [7]. The optimal values of the dissipation strength in the present work appear to be significantly smaller than in [7] due to somewhat higher barriers. The value of $\chi^2$ in the present RMF calculations are of the same order as in [7] not showing any regular trend.

TABLE II. Reactions for which results are presented in Fig. 4, the optimal values of the radial friction strength $K_{Rm}$, the corresponding minimum values of $\chi^2$. In the last two columns, the heights of Coulomb barrier at zero angular momentum $U_{B0}$ [with which the dynamical calculations are performed (NL2_P)] are compared with the ones obtained using the HS parametrization and the M3Y_P from Ref. [7].

| Reaction | $K_{Rm}$ (zs·GeV$^{-1}$) | | $\chi^2_m$ | | $U_{B0}$ (MeV) | |
|---|---|---|---|---|---|---|
| | this work, NL2_P | Ref. [7], M3Y_P | this work, NL2_P | Ref. [7], M3Y_P | this work, NL2_P HS_R HS_P | Ref. [7], M3Y_P |
| $^{16}$O+$^{92}$Zr | 10 | 27 | 6 | 17 | 42.1 42.71 41.90 | 41.6 |
| $^{16}$O+$^{144}$Sm | 3 | 16 | 5 | 8 | 61.4 62.18 61.05 | 60.7 |
| $^{16}$O+$^{208}$Pb | 6 | 13 | 71 | 69 | 76.3 77.26 75.97 | 75.6 |
| $^{28}$Si+$^{92}$Zr | 7 | 19 | 6 | 3 | 71.4 72.50 71.01 | 70.5 |
| $^{12}$C+$^{144}$Sm | 15 | 23 | 0.2 | 0.04 | 46.9 47.45 R 46.64 P | 46.4 |

The fusion excitation functions calculated using the NL2_P *NN*-forces with the optimal value $K_{Rm}$ are compared with the data in Fig. 4. Here the ratio of the cross-sections $\sigma_{th}/\sigma_{exp}$ is shown as the function of the ratio $U_{B0}/E_{c.m.}$ for five reactions listed in Table II. Typical statistical error of the Langevin modeling is about 1%. Fig. 4 indicates rather good quality of the theoretical description of the precision data: among 43 points only 3 are located beyond the 5% stripe around the unity.

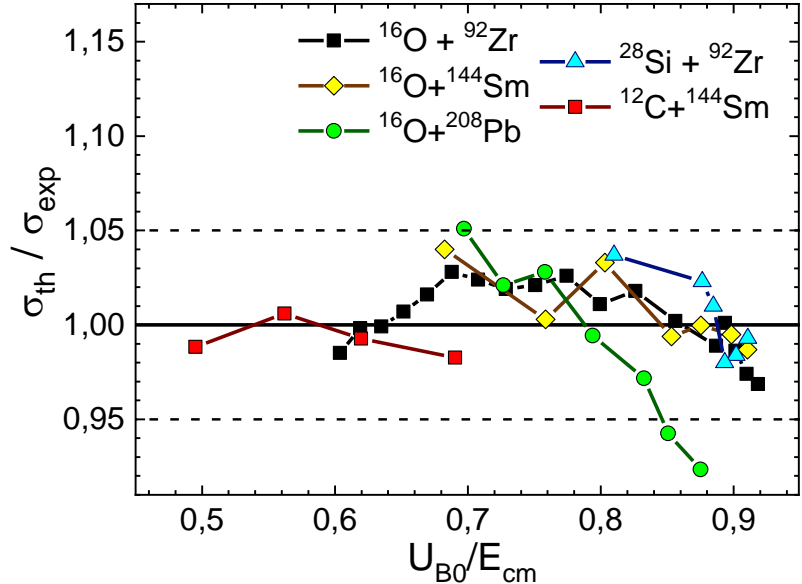

FIG. 4. The ratio $\sigma_{th}/\sigma_{exp}$ as the function of $U_{B0}/E_{c.m.}$ for five reactions.

To summarize, the above barrier portions of the fusion (capture) cross sections for heavy-ion induced reactions with spherical nuclei have been calculated using the nucleus-nucleus potential resulting from the relativistic mean field approach. The strong nucleus-nucleus potential has been calculated using the double folding model (see Eq. (3)). The relativistic mean field effective nucleon-nucleon (*NN*) forces v$_{NN}$ have been employed as the ingredients of the double-folding model. These *NN*-forces correspond to the NL1, NL2, NL3, and HS sets of the RMF parameters known in the literature. The nuclear matter densities came from the Skyrme-Hartree-Fock approach with the tensor forces with the SKX parametrization. These calculations prove that two of these parameter sets (NL1 and NL3) do not result in a total (strong nucleus-nucleus + Coulomb) potential with a barrier. Adding the delta-function term like in the M3Y Paris or Reid nucleon-nucleon forces results in the nucleus-nucleus potential with a barrier only in the case of NL2 parameter set (NL2_P) and HS-parameter set (HS_R). The reason is that these sets produce the nucleon-nucleon potentials $v_{NN}$ which increase similar to the M3Y forces as the distance between nucleons decreases. The relativistic mean field Coulomb barriers appear to be about 1% higher than those obtained using the M3Y *NN* forces.

The theoretical cross sections have been evaluated by means of the fluctuation-dissipation trajectory model with surface friction known in the literature, using the NL2_P forces. Results of the comparison point-by-point with the high precision experimental data (the typical error is about 1%) demonstrates good quality of the description: the typical value of $\chi^2$ is several units. Using the HS parameter set of Ref [11] is expected to lead to similar results.